\documentclass{elsart}
\usepackage{graphicx}
\usepackage{amsmath}
\usepackage{amssymb}

\begin{document}


\begin{frontmatter}

\title{Experimental study of a liquid Xenon PET prototype module}
\author{M.-L. Gallin-Martel\thanksref{corr}},
\author{P. Martin}, 
\author{F. Mayet},
\author{J. Ballon},
\author{G. Barbier},
\author{C. Barnoux},
\author{J. Berger},
\author{D. Bondoux},
\author{O. Bourrion},
\author{J. Collot},
\author{D. Dzahini},
\author{R. Foglio},
\author{L. Gallin-Martel},
\author{A. Garrigue},
\author{S. Jan\thanksref{shfj}},
\author{P. Petit},
\author{P. Stassi}, 
\author{F. Vezzu}, 
\author{E. Tournefier\thanksref{lapp}}

\thanks[corr]{Corresponding author : Tel.: +33-476-284-128; fax: +33-476-284-004; e-mail: mlgallin@lpsc.in2p3.fr.)}
\thanks[shfj]{Present address :  
Service Hospitalier Fr\'ed\'eric Joliot (SHFJ), CEA, F-91401 Orsay, France}
\thanks[lapp]{Present address : Laboratoire d'Annecy-le-Vieux de Physique des 
Particules, CNRS/IN2P3, BP 110, F-74941 Annecy-le-Vieux cedex, France}
\address{Laboratoire de Physique Subatomique et de Cosmologie, 
 CNRS/IN2P3 et Universit\'e Joseph Fourier, 
 53, avenue des Martyrs, F-38026 Grenoble cedex, France}

{\em \normalsize submitted to Proc. of the 7th International 
Workshops on Radiation Imaging Detectors, 4-7 july 2005, Grenoble, France} 

\begin{abstract}
A detector using liquid Xenon (LXe) in the scintillation mode is studied for Positron Emission Tomography (PET). The specific design aims at taking full advantage of the liquid Xenon properties. It does feature a promising solution insensitive to any parallax effect. This work reports on the spatial resolution capabilities of the first LXe prototype module, equipped with a Position Sensitive Photo-Multiplier Tube (PSPMT) operating in the VUV range (178 nm).
\end{abstract}


\begin{keyword}
Positron emission tomography (PET), Medical imaging equipment\\ 
{\it PACS : }87.58.Fg ; 87.62.+n
\end{keyword}
\end{frontmatter}


\section{Introduction}
Positron Emission Tomography (PET) is one of the leading techniques of nuclear medicine to access to metabolic and functional information. PET is used for various medical and biological applications, such as oncology, cardiology as well as pharmacology. 
Experimental efforts on a host of techniques have been made in the field of  PET  imaging, in particular towards the development of new generation high resolution PET cameras dedicated to small animal imaging \cite{weber,yang,wienhard}. A couple of years ago, we proposed to use liquid Xenon in an axial geometry for a scintillation based PET \cite{collot,itbs,thesejan}.
LXe shows rather attractive features when compared to commonly used crystals : the density 3 gcm$^{-3}$ is not so small for a liquid (close to the value of NaI), the decay time is short 3 - 30 ns, the range being due to the various scintillation modes of the Xe atom, the LXe light yield is very high (60 10$^3$ UV/MeV in average \cite{doke,seguinot}).

This paper reports on the performances of the first LXe prototype module, in terms of spatial resolution capabilities.

\section{The geometry of the Liquid Xenon PET camera}
The active part of this project of LXe camera is a ring featuring an internal diameter of 8 cm (see Fig. 1). It is filled with liquid Xenon and placed in a cryostat composed of thin aluminum walls. Sixteen identical modules of the type shown on Fig. 2, are immersed in this ring. Each module presents a $\rm 2 \times  2 \ cm^2$ cross-section in the transaxial plane of the camera. The axial field of view is 5 cm. A module is optically subdivided by one hundred $\rm 2 \times  2 \ mm^2$ $\rm MgF_2$-coated aluminum UV light guides. The UV light is collected on both sides of a module by two Position Sensitive Photo-Multiplier Tubes (PSPMT). The (x,y) positions measured by the photo-tubes determine which light guides have been fired. For each module, the axial coordinate (z) is provided by the following ratio of the PSPMT$_i$ (i=1,2) dynode signals :
$$
Z =(PSPMT_1-PSPMT_2)/(PSPMT_1+PSPMT_2)
$$

allowing to measure the three coordinates without any parallax error. 

\section{Experimental set-up}
Following the layout displayed on Fig. 3, the Xenon is liquefied in the compressor, then transferred to a container inside the cryostat. The temperature inside the cryostat is kept around 165 K via a liquid nitrogen heat exchanger. The temperature is constant to better than a few tenths of a degree. The Xenon container is a stainless steel cylinder 50 mm long and 40 mm in diameter, closed at each end with a suprasil 3 mm thick window. A $\rm ^{22}Na$ source  is mounted on a small carriage moving along the z direction. A LYSO crystal coupled to a photomultiplier tube completes the experimental set-up to make the coincidence signal. The VUV photons are then collected with one PSPMT at each end. PSPMT with the required specifications, i.e high Quantum Efficiency (QE) at 178 nm (QE = 20 $\%$) and still working at low temperatures (165 K), are not commercially available yet. Hamamatsu provided us with two prototype tubes, from the R8520-06-C12 series \cite{hamamatsu}, having five aluminum strips deposited on their window to improve the resistivity of the photocathode at 165 K.\\

\noindent
The read out electronics operates at room temperature and is composed of standard NIM and CAMAC modules. The data acquisition (DAQ in Fig. 3) software performs barycentre online calculation.

\section{Experimental results}
\subsection{PSMPT spatial resolution and x and y localization}
To evaluate the PSPMT spatial resolution, a deuterium light source is used to produce a constant number of photons in a wide wavelength range. To level down the number of photons emitted, dedicated light attenuators are placed in front of the light source. Then to focus the light pulse on a specific area defined on the PSPMT surface, the light emitted by the source is collected via an optical fibre going through an opaque plastic disk placed in front of the PSPMT window. This disk exhibits a matrix of holes equally spaced. The number of photoelectrons is derived from the dynode signal distribution. The x and y barycentre distributions are computed on-line by the DAQ and derived of the anode signals. The PSPMT resolution (FWHM) in x and y as a function of the number of photoelectrons ($\rm N_{pe}$) is illustrated by Fig. 4. The resolution is increasing with Npe, ranging from 0.32 mm down to 0.18 mm for $\rm N_{pe}$  $\geq$ 300. This experimental study concludes that the PSPMT spatial resolution is fine for our application. The resolution is at the level of 1 mm in the x and y directions which is in very good agreement with the simulation \cite{nim}.
The second step of this analysis was to study the light guide separation in the transaxial plane. It has been evaluated using an $^{241}$Am $\alpha$ source located at one end of the Xenon container. The advantage of using an a source in the liquid is to give intense point-like sources of photons, at a well defined distance of the PSPMT. Three different matrix of light guides were used for the tests, with cross sections of 2 $\times$ 2, 5 $\times$ 5 and 2 $\times$ 5 mm$^2$, within an overall cross section of $\rm 20 \times 20 \ mm^2$. In the 2 x 5 configuration, the module had therefore only 40 cells of 48 mm in length. The walls of these cells are made of a double $\rm 35 \ \mu m$ thick aluminum foil, double because specular reflection has been guarantied on one side only : the basic material is a $\rm 35 \ \mu m$ aluminum foil, with a thin polyethylene film glued on one side to reduce the crookedness of its surface, followed with evaporation of aluminum again and $\rm MgF_2$ to make the actual reflecting surface. The Fig. 5 obtained with a guide matrix of $\rm 2 \times 5 \ mm^2$ shows a very satisfying light guides separation in the (x,y) transverse plane (see Fig. 5).

\subsection{Localization and resolution along the z axis}
The specific design of the liquid Xenon PET prototype employs the Depth Of Interaction  (DOI) approach to solve the problem of parallax errors. It permits a continuous measurement of the z coordinate. An experimental test bench (see section 3 and Fig. 3) has been built to measure the module prototype resolution in z. The z coordinate is deduced from the amplitude of the dynode signals measured on the right and left PSPMT located at each side of the module, as defined in section 2.
The resolution at a z position will be given by the FWHM of the obtained distributions. The resolution as a function of the source localization is illustrated by Fig. 6. The resolution is better at the module extremities rather than at the central position where it is about 10 mm. This result is not as good as expected but a refined simulation of the light collection in optical guides has been done and three configuration for the light collection units (present setup with the PSPMT, PSPMT immersed in the LXe, windowless Avalanche Photo-Diode APD immersed in the LXe) have been compared \cite{nim}. The objective is to increase the amount of collected light. This analysis by simulation concludes that a higher resolution can be achieved by :
\begin{itemize}
\item increasing the reflectivity of the light guide using other manufacturing processes (the current value for the reflectivity is 0.78),
\item immersing the PSPMT in the liquid (not recommended by Hamamatsu),
\item using high quantum efficiency Avalanche Photodiodes (to be investigated soon).
\end{itemize}
\section{Conclusion}
First test of a liquid Xenon TEP prototype module were carried out. The preliminary results of the experimental study allow us to determine the intrinsic performance of this camera. The localization in the transaxial plane is very satisfying : the resolution of the PSPMT is better than 0.3 mm in the x and y directions. Efforts are to be made for the localization in the axial direction since the resolution in z is not only positon dependent but exhibits a poor average value of about 8 mm. Simulation of the light collection in optical guides and in various photodetector devices (PSPMT at the LXe temperature but not immersed in the liquid, PSPMT immersed in the LXe, windowless APD immersed in the LXe) concludes that a higher resolution can be achieved \cite{nim}. Two next steps are foreseen. At first, the light guide reflectivity (currently 0.78) is to be improved, a new process is under study. Then the VUV light collection on each module end would be better by using high quantum efficiency windowless APD. We aim to investigate now these two possibilities.

\noindent \textbf{Acknowledgments : }\\
This work has been made possible thanks to the financial grants allocated by the Rh\^one-Alpes region through its ``Emergence'' science program, and  by the CNRS/INSERM via its IPA program dedicated to the small animal imaging. We are also indebted to Jean-Fran\c{c}ois Le Bas and Daniel Fagret of the medical department of the Joseph Fourier University of Grenoble for the support and motivation they brought to this project. We also wish to thank the technical staff of LPSC and in particular : Y. Carcagno, P. Cavalli, E. Lagorio, G. Mondin, A. Patti and E. Perbet.



\newpage

\begin{figure}[p]
\begin{center}
\includegraphics[scale=0.7]{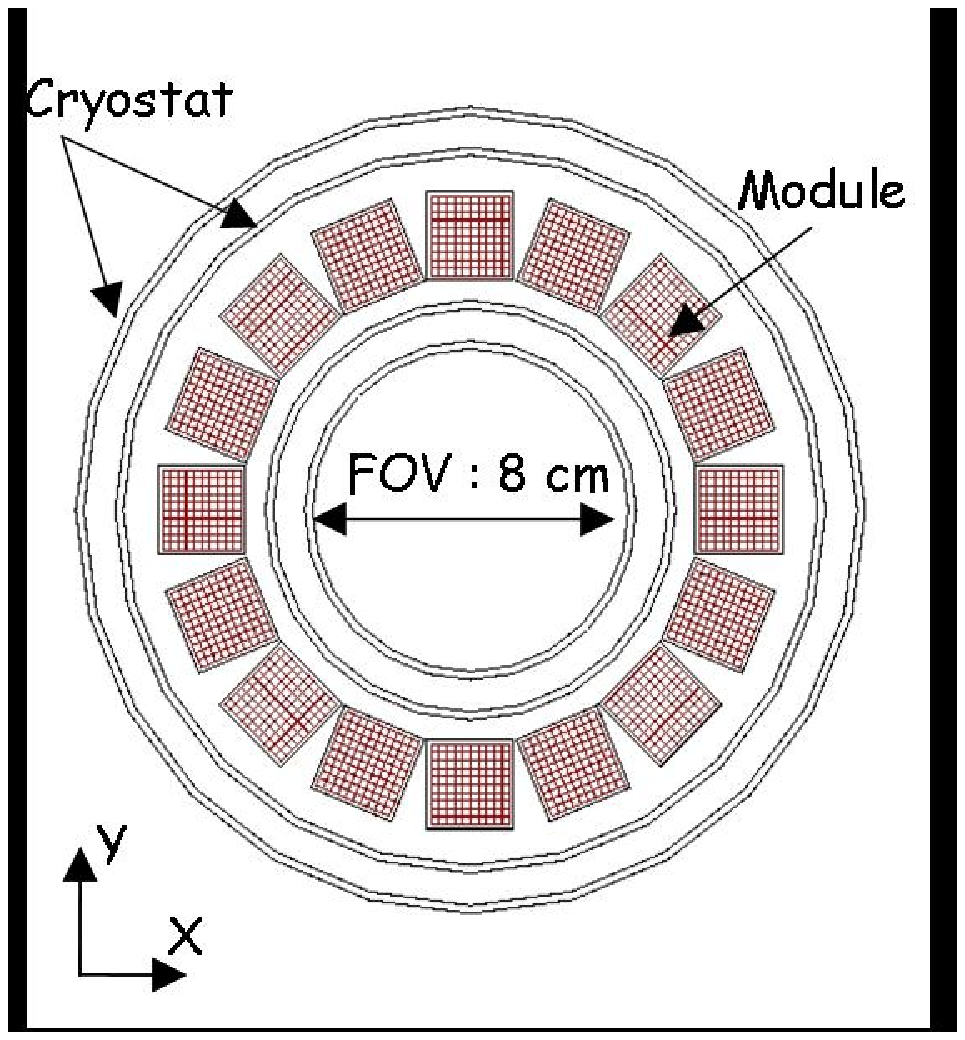}   
\caption{Transaxial view of the LXe $\mu$PET}
\label{fig:lxe}
\end{center}
\end{figure}

\begin{figure}[p]
\begin{center}
\includegraphics[scale=0.5]{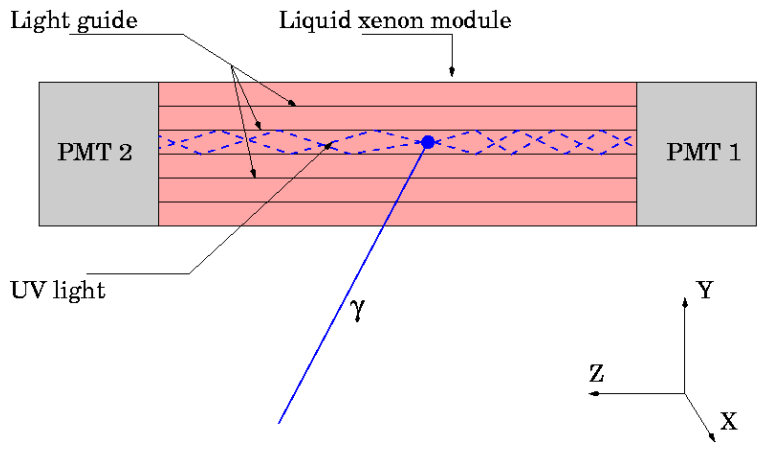}   
\caption{Schematic of an elementary module of the LXe camera : 
the module dimension are 2 $\times$ 2 $\times$  5 cm$^3$.}
\label{fig:Module}
\end{center}
\end{figure}

\begin{figure}[p]
\begin{center}
\includegraphics[scale=0.6]{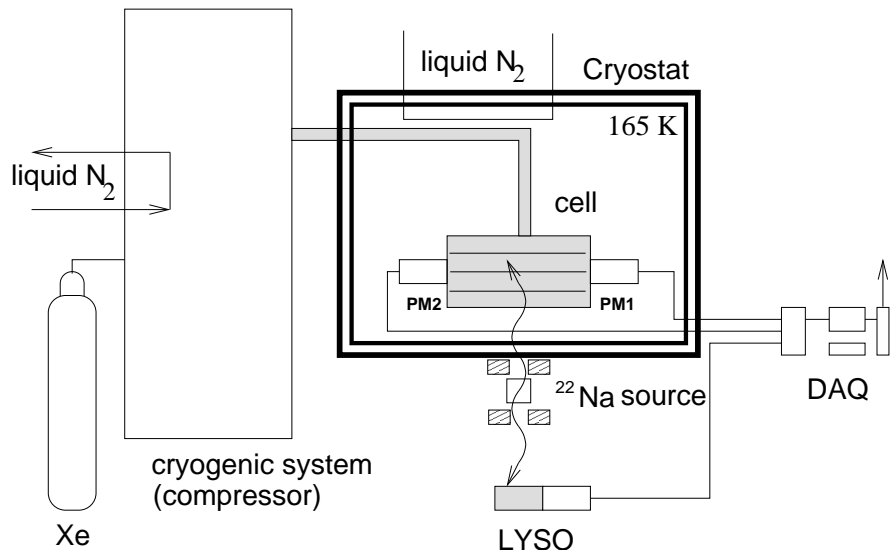}
\caption{The experimental set-up.}
\label{figure:exp}
\end{center}
\end{figure}

\newpage

\begin{figure}[p]
\begin{center}
\hspace*{-10mm}\includegraphics[scale=0.8,angle=0]{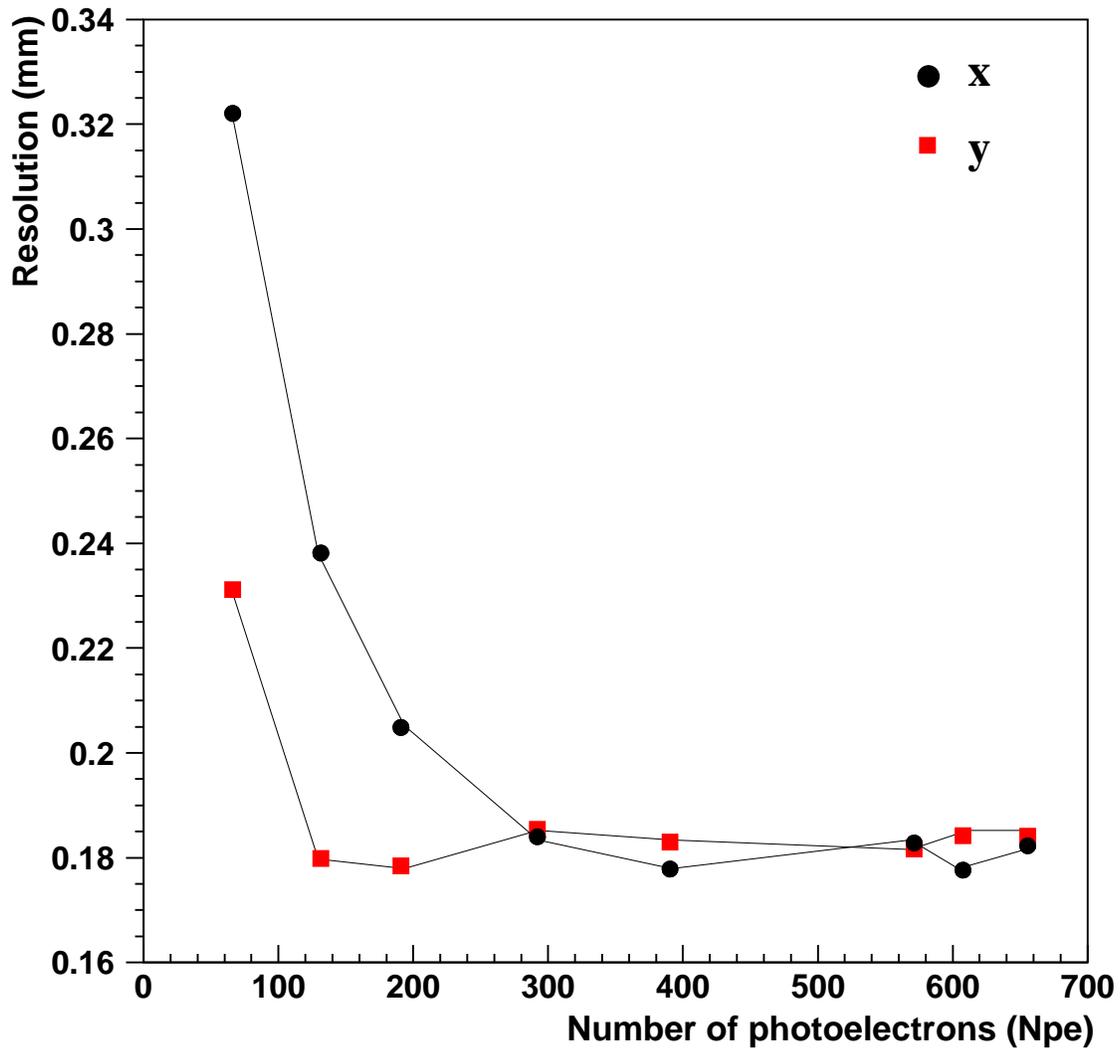}
\caption{The PSPMT resolution (FWHM) in X and Y as a function of the number of
photoelectrons.}
\label{fig:res.3}
\end{center}
\end{figure}

\begin{figure}[p]
\begin{center}
\includegraphics[scale=3.5,angle=0.]{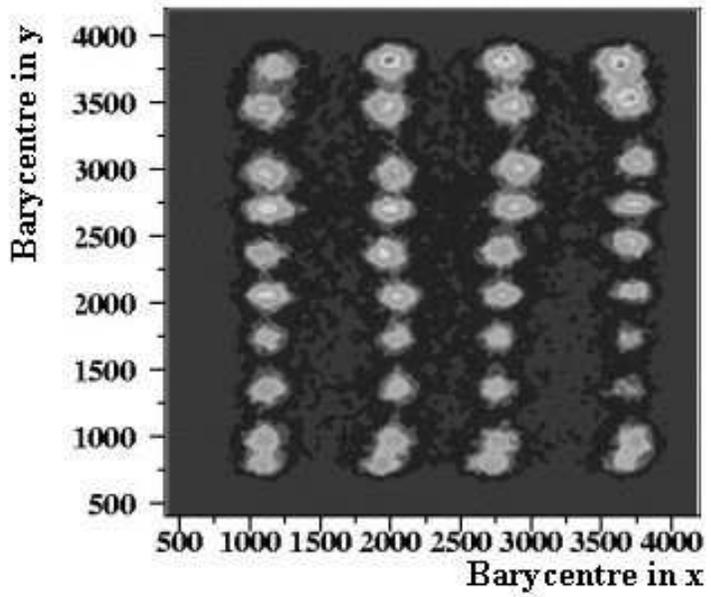}
\caption{X and Y localization using the 2 $\times$ 5 mm$^2$ light guides configuration.}
\label{fig:res.2}
\end{center}
\end{figure}

\begin{figure}[h]
\begin{center}
\includegraphics[scale=0.35,angle=0]{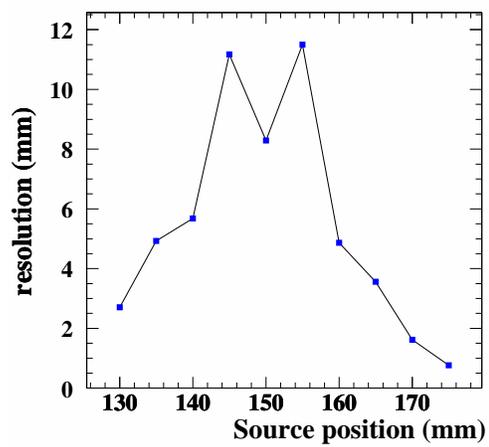}
\caption{Axial resolution as a function of the source localization}
\label{fig:res.1}
\end{center}
\end{figure}

\end{document}